\documentclass[conference,10pt,twocolumn,twoside]{IEEEtran}
\IEEEoverridecommandlockouts
\usepackage{amsmath}
\usepackage{amsfonts}
\usepackage{cases}
\usepackage{setspace}
\usepackage{fancybox}
\usepackage{subfigure}
\usepackage{epsfig}
\usepackage{graphicx}
\usepackage{epstopdf}
\usepackage{float}
\usepackage{multirow}
\usepackage{color}
\usepackage{amsmath}
\usepackage{multirow}
\usepackage{indentfirst}
\usepackage{dsfont}
\usepackage{amsfonts}
\usepackage{times,amsmath,color,amssymb,epsfig,cite,subfigure,algorithm,algorithmic}
\usepackage{todonotes}

\newcommand{\qed}{\nobreak \ifvmode \relax \else
      \ifdim\lastskip<1.5em \hskip-\lastskip
      \hskip1.5em plus0em minus0.5em \fi \nobreak
      \vrule height0.40em width0.6em depth0.25em\fi}

\newtheorem{lemma}{Lemma}
\newtheorem{proposition}{Proposition}
\newtheorem{theorem}{Theorem}

\begin{document}

\title{Unimodular Waveform Design for Integrated Sensing and Communication MIMO System via Manifold Optimization}

\author{
\IEEEauthorblockN{Jiangtao Wang\textsuperscript{1}, Xuyang Zhao\textsuperscript{2}, Muyu Mei\textsuperscript{3}, and Yongchao Wang\textsuperscript{1,}\textsuperscript{2}}

\IEEEauthorblockA{\textsuperscript{1}Guangzhou Institute of Technology, Xidian University, China}
\IEEEauthorblockA{\textsuperscript{2}State Key Laboratory on ISN, Xidian University, China}
\IEEEauthorblockA{\textsuperscript{3}School of Computer Science and Communication Engineering,
Jiangsu University, China}
Email: jtwang@xidian.edu.cn

}

\maketitle

\begin{abstract}
  Integrated sensing and communication (ISAC) has been widely recognized as one of the key technologies for 6G wireless networks.
  In this paper, we focus on the waveform design of ISAC system, which can realize radar sensing while also facilitate information transmission.
  The main content is as follows: first, we formulate the waveform design problem as a nonconvex and non-smooth model with a unimodulus constraint based on the measurement metric of the radar and communication system.
 Second, we transform the model into an unconstrained problem on the Riemannian manifold and construct the corresponding operators by analyzing the unimodulus constraint.
  Third, to achieve the solution efficiently, we propose a low-complexity non-smooth unimodulus manifold gradient descent (N-UMGD) algorithm with theoretical convergence guarantee.
   The simulation results show that the proposed algorithm can concentrate the energy of the sensing signal in the desired direction and realize information transmission with a low bit error rate.
\end{abstract}


\IEEEpeerreviewmaketitle

\section{Introduction}

\IEEEPARstart{R}{\MakeLowercase a}dar sensing and communication are widely used in the field of radio technology.
Rapid growth of wireless services leads to a shortage of spectrum resources, leading to co-frequency interference between communication and radar sensing systems \cite{Shlezinger_20}.
However, the development of information technology and equipment integration technology make it possible to integrate sensing and communication (ISAC) system.

ISAC has many technical advantages, simplifying system structure, reducing overhead, suppressing the co-frequency interference, and so on.
As a result, many researchers have been attracted to the joint design of sensing and communication functionalities \cite{Zhou_22}.
Among them, waveform design determines the performance boundary that the system can achieve and is one of the most popular topics \cite{Zhang_21}.

The waveform design of ISAC system can be roughly divided into three stages: in the first stage, the two systems share the same spectrum resource and achieve their own functions \cite{Hassanien_16},\cite{Liu_18_twc}.
The goal at this stage of waveform design is to avoid mutual interference as much as possible \cite{Shi_18},\cite{Liu_18}.

The second stage mainly adopts a kind of loose coupling technology, and the sensing and communication systems are integrated and mutually assisting.
Hardware devices are shared based on the multiplexed resources such as time, frequency, space or code \cite{Cao_20}-\hspace{-0.001cm}\cite{He_24}.
Design ideas often focus on the basis of communication or sensing, and achieve the other function under a certain constraint \cite{Keskin_21},\cite{Bazzi_23}.

Under the quality of service requirements of the communication and constant-modulus power constraint, the work in \cite{Liu_21_jstsp} focused on transmit beamforming designs. Different from the traditional work, this paper presents an efficient algorithm for solving nonconvex problems on Riemannian space.

The third stage is the tightly coupled technology stage, in which the sensing and communication systems are closely integrated.
The design goal is no longer to improve the sensing performance or communication performance alone, but to improve the overall performance, focusing on solving the diverse needs of future sensing and communication \cite{Xiao_22},\cite {Zhang_24}.
In \cite{Hassanien_16_tsp}, transmit weight vector was designed to maximize the signal energy of the desired direction, and control the amount of transmit power radiated towards the communication directions, so as to realize the dual functions of radar sensing and communication.

The Pareto optimal model of Cramer-Rao Bound and distortion minimum mean square error metric was formulated in \cite{Kumari_20} to quantify the trade-off between radar sensing and communication.

Most of the above literatures are iterative algorithms on the Euclidean space, using approximate relaxation technique to deal with the nonconvex unimodulus constraint, resulting in the algorithm either cannot prove convergence, or the performance has a certain gap from the theoretical boundary.

In this paper, we focus on the design of unimodular waveform for ISAC system.
The paper's main content is as follows:

first, the performance metrics for measuring radar sensing and communication are discussed, and a nonconvex nonsmooth optimization problem with unimodulus constraint is formulated.

Second, we employ Riemannian geometry theory to analyze the inherent geometry of unimodulus constraint and construct Riemannian manifold and its operators.

Third, we propose a customized nonsmooth unimodulus manifold gradient descend (N-UMGD) algorithm and analyze its convergence.

Simulation results demonstrate that the designed waveform can achieve excellent performance in radar sensing and communication.

\section{System Model and Problem Formulation}
\begin{figure}[t]
\centerline{\includegraphics[scale=0.6]{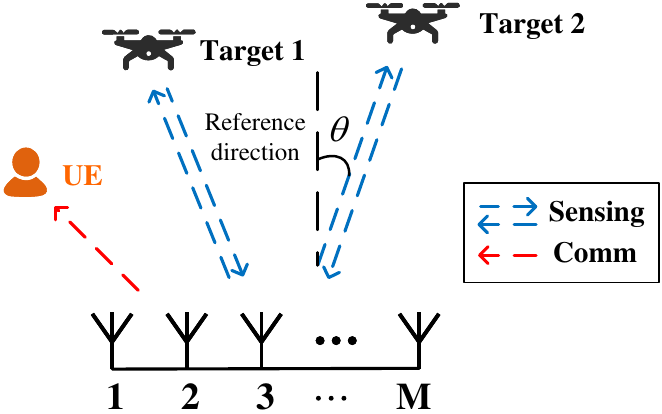}}
\caption{ISAC system working scenario with multiple targets/UE.}
\label{model}
\end{figure}
We consider an ISAC system with $M$-antenna uniform linear array (ULA) shown in figure \ref{model}.
The waveform transmitted by the $m$-th antenna can be expressed as
\begin{equation}\label{def x}
\begin{split}
\mathbf{x}_m = \left[x_{m}(1),x_{m}(2),\cdots,x_{M}(N)   \right]^T.
\end{split}
\end{equation}
Then, the waveform matrix can be denoted as $\mathbf{X}=[\mathbf{x}_1,\mathbf{x}_2,\cdots,\mathbf{x}_M]$.
Let $\mathbf{x}_n=[x_1(n),x_2(n),\cdots,x_M(n)]^T$ represent the $n$-th sample vector of $M$ transmitted waveforms.
The aim of this paper is to design the waveform $\mathbf{X}$, which can realize multiple targets sensing and communication.
Therefore, the metrics for measuring the quality of sensing and communication are discussed separately below.

\subsection{Radar Sensing Performance Metrics}
The steering vector at direction $\theta$ is defined as follows
\begin{equation}\label{steering vector}
\begin{split}
\mathbf{a}_{\theta}=\left[ 1,e^{j2\pi\frac{d}{\lambda}sin\theta},\cdots,e^{j2\pi\frac{d}{\lambda}(M-1)sin\theta}\right]^T,
\end{split}
\end{equation}
where $d$ is the array spacing and $\theta$ belongs to angle set $\Theta = \left(-90^{\circ}, 90^{\circ}\right)$.
The corresponding synthesized signal is given by
\begin{equation}\label{def syn}
\begin{split}
s_{\theta,n}=\mathbf{a}_{\theta}^T\mathbf{x}_n.
\end{split}
\end{equation}

The beampattern  which describes the radiation intensity distribution of the antenna array at direction $\theta$ can be written as
\begin{equation}\label{def beam}
\begin{split}
P_{\theta}(\mathbf{X})=\sum_{n=1}^Ns_{\theta,n}^*s_{\theta,n}=\mathbf{a}_{\theta}^H\mathbf{X}^H\mathbf{X}\mathbf{a}_{\theta}.
\end{split}
\end{equation}
Based on this, the spacial correlation characteristics of transmit waveform at different directions can be described as
\begin{equation}\label{def spa cor}
\begin{split}
P_{\theta_{ij},\tau}(\mathbf{X})=\mathbf{a}_{\theta_i}^H\mathbf{X}^H\mathbf{S}_\tau\mathbf{X}\mathbf{a}_{\theta_j},
\end{split}
\end{equation}
where the parameter $\tau$  represents the waveform relative delay and $\mathbf{S}_\tau \in \mathds{R}^{N\times N}$ is the shift matrix given by
\[
\begin{split}
&{\mathbf{S}_\tau}(i,j) =
\left\{
  \begin{array}{l}
  1, \ {\rm if }\ j -i =\tau,\\
  0, \ {\rm  otherwise,}
  \end{array}
\right.
\end{split}
\]
where $i,j\in\{1,\cdots,N\}$.

This paper consider the following metrics in the design of radar sensing waveform.
\begin{itemize}
\item Desired beampattern design
\end{itemize}

In order to obtain greater antenna gain, higher system capacity and longer transmission distance, the energy of the transmitted waveform is concentrated in the desired target direction and the clutter energy in other directions is suppressed.

The beampattern design model can be expressed as follows
\begin{equation}\label{diff sum}
\begin{split}
B(\mathbf{X})=\sum_{\theta\in\bar{\Theta}}\mathbf{a}_{\theta}^H\mathbf{X}^H\mathbf{X}\mathbf{a}_{\theta}-\sum_{\theta\in\hat{\Theta}}\mathbf{a}_{\theta}^H\mathbf{X}^H\mathbf{X}\mathbf{a}_{\theta}.
\end{split}
\end{equation}
where $\bar{\Theta}=\{\theta_1,\cdots,\theta_K\}$ and $\hat{\Theta} \in \Theta - \bar{\Theta}$ are the angle sets of desired and unconsidered direction, respectively.

\begin{itemize}
\item Spacial correlation sidelobe levels
\end{itemize}

Based on \eqref{def spa cor}, the matrix used to describe the spacial sidelobe levels is defined by
\begin{equation}\label{spa corr}
\begin{split}
  &\mathbf{C}(\mathbf{X})\!\! =\!\!\!
  \begin{bmatrix}
     P_{\theta_{11},\tau_1}(\mathbf{X})  & P_{\theta_{11},\tau_2}(\mathbf{X}) & \dotsb &P_{\theta_{11},|\Gamma|}(\mathbf{X}) \\
     P_{\theta_{12},\tau_1}(\mathbf{X})  & P_{\theta_{12},\tau_2}(\mathbf{X}) & \dotsb &P_{\theta_{12},|\Gamma|}(\mathbf{X}) \\
      \vdots & \vdots & \vdots & \vdots & \\
     P_{\theta_{ij},\tau_1}(\mathbf{X})  & P_{\theta_{ij},\tau_2}(\mathbf{X}) & \dotsb &P_{\theta_{ij},|\Gamma|}(\mathbf{X}) \\
           \vdots & \vdots & \vdots & \vdots & \\
     P_{\theta_{KK},\tau_1}(\mathbf{X})  & P_{\theta_{KK},\tau_2}(\mathbf{X}) & \dotsb &P_{\theta_{KK},|\Gamma|}(\mathbf{X}) \\
  \end{bmatrix},
\end{split}
\end{equation}

where $\Gamma$ represents a set of delay parameters $\{\tau_1,\tau_2,\cdots\in \Gamma\}$.
Minimizing spacial correlation sidelobe levels at desired direction can improve spacial resolution and reduce the influence of clutter components generated by other targets.

\subsection{Signal Transmission Mechanism and Communication Performance Metric}
The goal of this paper is that the designed waveform can achieve radar sensing while also facilitate information transmission. Specifically, the transmitter encodes the input information block into codewords with symbols drawn from alphabet $\{-a, a\}$ and represents the symbols $a$ and $-a$ by sending waveforms $\mathbf{-X}$ and $\mathbf{X}$, respectively. 

The desired communication user at direction $\theta_c$ aims to determine whether the transmitted waveform is $\mathbf{-X}$ or $\mathbf{X}$ through signal detection. Mathematically, this can be represented as a binary hypothesis testing problem. One hypothesis, denotes as $\mathcal{H}_0$, assumes that the transmitter is transmitting waveform $\mathbf{-X}$. The other hypothesis, denotes as $\mathcal{H}_1$, assumes that the transmitting waveform is $\mathbf{X}$. Therefore, the signal detection problem is as follows
\begin{equation}\label{signal-detection-problem}
\begin{array}{ll}
\mathcal{H}_0: \mathbf{y}=-\mathbf{X}\mathbf{a}_{\theta_c}+\mathbf{n},  \\
\mathcal{H}_1: \mathbf{y}=\mathbf{X}\mathbf{a}_{\theta_c}+\mathbf{n},
\end{array}
\end{equation}
where $\mathbf{n} \sim \mathcal{C N}\left(0, \sigma^2\mathbf{I}_N\right)$ denotes the AWGN at the desired communication user.

Let $p_0\left(\mathbf{y}\right)$ and $p_1\left(\mathbf{y}\right)$ denote the likelihood functions of the received signals of the desired user under $\mathcal{H}_0$ and $\mathcal{H}_1$, respectively. Based on \eqref{signal-detection-problem},  $p_0\left(\mathbf{y}\right)$ and $p_1\left(\mathbf{y}\right)$ are given as
\begin{subequations}
\begin{align}
&p_0(\mathbf{y})\!=\! \frac{1}{\pi^{n} \operatorname{det}\mathbf{C}} \exp \left[(\mathbf{y}\!+\!\mathbf{X}\mathbf{a}_{\theta_c})^H \mathbf{C}^{-1}(\mathbf{y}\!+\!\mathbf{X}\mathbf{a}_{\theta_c})\right],\\
&p_1(\mathbf{y})\!=\! \frac{1}{\pi^{n} \operatorname{det}\mathbf{C}} \exp \left[(\mathbf{y}\!-\!\mathbf{X}\mathbf{a}_{\theta_c})^H \mathbf{C}^{-1}(\mathbf{y}\!-\!\mathbf{X}\mathbf{a}_{\theta_c})\right],
\end{align}
\end{subequations}
where $\mathbf{C} =\sigma^2\mathbf{I}_N$.
The bit error rate (BER) of the system can be expressed as
\begin{equation}\label{def ber}
\mathbb{P}_e = \mathbb{P}\left(\mathcal{H}_0|\mathcal{H}_1\right)\mathbb{P}\left(\mathcal{H}_1\right)+ \mathbb{P}\left(\mathcal{H}_1|\mathcal{H}_0\right)\mathbb{P}\left(\mathcal{H}_0\right).
\end{equation}

The desired user employs a maximum likelihood (ML) detector. Since the transmitter sends $-\mathbf{X}$ and $\mathbf{X}$ equiprobable, we have $\mathbb{P}\left(\mathcal{H}_0\right)=\mathbb{P}\left(\mathcal{H}_1\right)=0.5$. Thus, the decision criterion is as follows\cite{Statistical_signal_processing}
\begin{equation}
\frac{p_1\left(\mathbf{y}\right)}{p_0\left(\mathbf{y}\right)}\mathop{\gtrless}\limits_{\mathcal{H}_0}^{\mathcal{H}_1}1.
\end{equation}

Then we can obtain
\begin{subequations}
    \begin{align}
       & \mathbb{P}\left(\mathcal{H}_1|\mathcal{H}_0\right) = Q\left(\frac{\|\mathbf{X}\mathbf{a}_{\theta_c}\|_2^2}{\sqrt{\|\mathbf{X}\mathbf{a}_{\theta_c}\|_2^2\sigma^2}}\right),\\
       & \mathbb{P}\left(\mathcal{H}_0|\mathcal{H}_1\right) = 1 - Q\left(\frac{-\|\mathbf{X}\mathbf{a}_{\theta_c}\|_2^2}{\sqrt{\|\mathbf{X}\mathbf{a}_{\theta_c}\|_2^2\sigma^2}}\right).
    \end{align}
\end{subequations}
Then, we bring them into \eqref{def ber} and obtain the specific expression for BER
\begin{equation}\label{def pe}
    \mathbb{P}_e(\mathbf{X}) \!= Q\left(\!\frac{\|\mathbf{X}\mathbf{a}_{\theta_c}\|_2^2}{\sqrt{\|\mathbf{X}\mathbf{a}_{\theta_c}\|_2^2\sigma^2}}\!\right).
\end{equation}

\subsection{Problem Formulation and Reformulation}

In order to  maximize the efficiency of the power amplifier,  the waveform unimodulus constraint is considered.
Therefore, based on the metrics discussed above for measuring sensing and communication, the waveform design problem for ISAC system is formulated as
\begin{equation}\label{ori pro}
\begin{split}
\hspace{-0.2cm}&\min_{\mathbf{X} \in \mathbb{C}^{N \times M}}\ \ \ -B(\mathbf{X})+\|\mathbf{c}(\mathbf{X})\|_1+\mathbb{P}_e(\mathbf{X}), \\
\hspace{-0.2cm}&{\rm subject\ to}\ |x_m(n)|\!=\!1,m\!=\!1,\cdots,M,n\!=\!1,\cdots,N,
\end{split}
\end{equation}
where $\mathbf{c}(\mathbf{X}) = \rm{vec}(\mathbf{C}(\mathbf{X}))$.

{\it \bf Remark:}

Compared to classical $\ell_2$-norm models, $\ell_1$-norm model proposed in \eqref{ori pro} will tend to obtain a large number of zero terms and a small number of nonzero terms.
However, this kind of modeling will also cause the objective function to be nonsmooth, which poses a challenge for the design of subsequent solving algorithms.

Since $Q$ function in \eqref{def pe} is strictly monotonically decreasing, minimizing $\mathbb{P}_e$ can be equivalently transformed into maximizing $\|\mathbf{X}\mathbf{a}_{\theta_c}\|_2^2$.
However, the objective function is still a nonconvex  polynomial, and the unimodulus constraint is difficult to deal with directly.
To effectively handle the unimodulus constraint, we employ Riemannian geometry theory to analyze the inherent geometric structure of the constraint and embed it into the search space. Then the original optimization problem can be converted into the following unconstrained form on the manifold $\operatorname{UM}\left(N,M\right)$.
\begin{equation}
\min_{\mathbf{X} \in \operatorname{UM}\left(N, M\right)}\ \ \ g\left(\mathbf{X}\right) + h\left(\mathbf{X}\right),
\end{equation}
where
\begin{subequations}
\begin{align}
g\left(\mathbf{X}\right) &=\sum_{\theta\in\hat{\Theta}}\mathbf{a}_{\theta}^H\mathbf{X}^H\mathbf{X}\mathbf{a}_{\theta}-\sum_{\theta\in\bar{\Theta}\cup \theta_c}\mathbf{a}_{\theta}^H\mathbf{X}^H\mathbf{X}\mathbf{a}_{\theta},\label{gx}\\
h\left(\mathbf{X}\right)&=\|\mathbf{c}(\mathbf{X})\|_1.\label{hx}
\end{align}
\end{subequations}

The analysis of the geometric properties of the manifold $\operatorname{UM}\left(N, M\right)$ and its corresponding operators are discussed in the next section.

\section{Constraint Analysis}

In this section, we first analyze the geometric properties of unimodulus constraints and then construct UM$(N,M)$ manifold.
In order to solve the optimization problem on UM$(N,M)$ manifold, it is necessary to generalize the concepts of direction and inner product from Riemannian space, and then construct operators on the manifold: Riemannian gradient and retraction.

\subsection{UM Manifold Structure Analysis}
For any complex vector $\mathbf{x} \in \mathbb{C}^{N \times 1}$, we have
\begin{equation}\label{real-representation}
\phi(\mathbf{x})=\left(\begin{array}{cc}
\operatorname{Re}(\mathbf{x}) & \operatorname{Im}(\mathbf{x}) \\
-\operatorname{Im}(\mathbf{x}) & \operatorname{Re}(\mathbf{x})
\end{array}\right).
\end{equation}

The complex vector space $\mathbb{C}^{N \times 1}$ is isomorphic to the quasi-symplectic set $\mathcal{SP}\left(1, N\right)$ as follows
\begin{equation}
\mathcal{S P}(1, N)=\left\{\tilde{\mathbf{X}} \in \mathbb{R}^{2 N \times 2 } \mid \tilde{\mathbf{X}} \mathbf{J}_1=\mathbf{J}_N \tilde{\mathbf{X}}\right\},
\end{equation}
where $\mathbf{J}_N=\left(\begin{array}{cc}0 & \mathbf{I}_N \\ -\mathbf{I}_N & 0\end{array}\right)$.

Then we use $\tilde{\mathbf{X}}$ to represent a matrix $\phi\left(\mathbf{x}\right)$, where $\mathbf{x} \in \operatorname{UM}\left(N , 1\right)$ and $\operatorname{UM}\left(N, 1\right) = \{\mathbf{x} \in \mathbb{C}^{N \times 1} \big{|}\ |x_i| = 1, i = 1,...,N\}$. We can find that all the rows of $\tilde{\mathbf{X}}$ have unit Euclidean norm. Let $\mathbf{Y} = \tilde{\mathbf{X}}^T$, we can conclude that the set of matrix $\mathbf{Y}$ satisfies
\begin{equation}
    \mathbf{Y} \in \bar{\mathcal{M}}:= \mathcal{O B}(2, 2N) \cap \mathcal{S P}(N, 1),
\end{equation}
where $\mathcal{OB}\left(2, 2N\right)$ is oblique manifold, i.e.,
\begin{equation}
\mathcal{O B}(2, 2N)=\left\{\mathbf{Y} \in \mathbb{R}^{2 \times 2N}: \operatorname{ddiag}\left(\mathbf{Y}^T \mathbf{Y}\right)=\mathbf{I}_N\right\}.
\end{equation}

According to the inherent property $\phi\left(\mathbf{x}\right)^T \in \mathcal{SP}\left(N, 1\right)$, for any $\mathbf{x} \in \mathbb{C}^{N\times 1}$ that satisfies $|x_i|=1$, we can transform it into the corresponding $\mathbf{Y} = \phi\left(\mathbf{x}\right)^T$ processing. At this point, $\mathbf{Y} \in \bar{\mathcal{M}}$ can be simplified to $\mathbf{Y} \in \mathcal{OB}\left(2, 2N\right)$.

Then, we can derive the properties of $\mathbf{Y}$ based on some well-known properties of $\mathcal{OB}\left(2, 2N\right)$.
According to \cite{Oblique_manifold}, the tangent space to $\mathcal{OB}\left(2, 2N\right)$ at $\mathbf{Y}$ is
\begin{equation}
T_{\mathbf{Y}} \mathcal{O B}\left(2, 2N\right)=\left\{\mathbf{Z}: \operatorname{ddiag}\left(\mathbf{Y}^T \mathbf{Z}\right)=\mathbf{0}\right\}.
\end{equation}

Then, the tangent space to $\bar{\mathcal{M}}$ at $\mathbf{Y}$ is
\begin{equation}\label{tangent_space_M}
    T_{\mathbf{Y}}\bar{\mathcal{M}} = \{\bar{\xi} \in \mathcal{SP}\left(N, 1\right)| \operatorname{ddiag}\left(\mathbf{Y}^T \mathbf{\xi}\right)=\mathbf{0}\},
\end{equation}
which means that $\mathbf{y}_i^T\mathbf{z}_i=0, i=1,...,N$, where $\mathbf{y}_i$ denotes the $i$-th column of $\mathbf{Y}$.

When considering $\mathcal{SP}\left(N, 1\right)$ as a subspace of $\mathbb{R}^{2 \times 2N}$, the inner product in $\mathcal{SP}\left(N, 1\right)$ can be expressed as

\begin{equation}\label{Sp-inner-product}
\langle\mathbf{Y}_1, \mathbf{Y}_2\rangle=\frac{1}{2} \operatorname{tr}\left(\mathbf{Y}_1^T \mathbf{Y}_2\right) , \quad \mathbf{Y}_1, \mathbf{Y}_2\in \mathcal{S P}(N, 1).
\end{equation}

Additionally, we can obtain the orthogonal projection operator on the tangent space $T_{\mathbf{Y}}\tilde{\mathcal{M}}$ of $\tilde{\mathcal{M}}$.
\begin{proposition}\label{real_projection}
 For any $\tilde{\xi}\in \mathcal{S} \mathcal{P}(N, 1)$, the orthogonal projection operator $P_{\mathbf{Y}}$ onto the tangent space $T_{\mathbf{Y}}\bar{\mathcal{M}}$ at $\mathbf{Y}$ is given by
\begin{equation}\label{M_projection}
    \operatorname{P}_{\mathbf{Y}}\tilde{\xi} = \tilde{\xi}-\mathbf{Y}\operatorname{ddiag}\left(\mathbf{Y}^T\tilde{\xi}\right).
\end{equation}
\end{proposition}

Then, we transform the mapping in {\it Proposition \ref{real_projection}} into complex vector operations and derive the expression for the orthogonal projection operator onto $T_{\mathbf{x}}\operatorname{UM}\left(N, 1\right)$ based on \eqref{M_projection}, resulting in the following lemma.
\begin{lemma}\label{projection-vector}
    For any $\xi \in \mathbb{C}^{N \times 1}$, the orthogonal projection operator $\operatorname{P}_{\mathbf{x}}$ onto the tangent space $T_{\mathbf{x}}\operatorname{UM}\left(N, 1\right)$ at $\mathbf{x}$ is given by
    \begin{equation}
       \operatorname{P}_{\mathbf{x}}\xi =  \xi - \operatorname{Re}\left(\xi \circ \mathbf{x}^*\right)\circ \mathbf{x},
    \end{equation}
\end{lemma}
where $\mathbf{x} \circ \mathbf{y}$ represents the Hademar product of $\mathbf{x}$ and $\mathbf{y}$.

\begin{lemma}\label{projection-matrix}
For any $\boldsymbol{\xi} \in \mathbb{C}^{M \times N}$, $\mathbf{X} \in \operatorname{UM}\left(N, M\right)$, the orthogonal projection operator $\operatorname{P}_{\mathbf{X}}$ onto the tangent space $T_{\mathbf{X}}\operatorname{UM}\left(N, M\right)$ at $\mathbf{X}$ is given by
\begin{equation}\label{[projection-gradient]}
    \operatorname{P}_{\mathbf{X}}\boldsymbol{\xi} = \boldsymbol{\xi} - \operatorname{Re}\left(\boldsymbol{\xi} \circ \mathbf{X}^*\right)\circ \mathbf{X}.
\end{equation}
\end{lemma}
{\it \bf Remark:} It is easy to obtain that $\operatorname{UM}\left(N, M\right)$ is the Cartesian product of $\operatorname{UM}\left(N, 1\right)$, i.e.,$\operatorname{UM}\left(N, M\right) = \operatorname{UM}\left(N, 1\right) \times ... \times \operatorname{UM}\left(N, 1\right) = \left(\operatorname{UM}\left(N, 1\right)\right)^M$. According to \cite{manifold-nicolas}, the tangent space of $\operatorname{UM}\left(N, M\right)$ at $\mathbf{X}$ is the Cartesian product of $T_{\mathbf{x}}\operatorname{UM}\left(N, 1\right)$, i.e., $T_{\mathbf{X}}\operatorname{UM}\left(N, M\right) = \left(T_{\mathbf{x}}\operatorname{UM}\left(N, 1\right)\right)^{M}$. Therefore, the conclusion of {\it Lemma \ref{projection-matrix}} can be easily derived from {\it Lemma \ref{projection-vector}}.

\subsection{Operator Construction of UM Manifold}
Consider a Riemannian optimization method on a manifold, we first find a steepest decrease direction, which is the negative gradient direction $-\operatorname{grad}f\left(\mathbf{X}\right)$. In each iteration, we perform a search starting from point $\mathbf{X}$ in the direction of $-\operatorname{grad}f\left(\mathbf{X}\right)$. However, this operation is too complicated, so we introduce retraction to ensure that the update point remains on the manifold after each update.

{\it Riemannian Gradient:} Given a smooth real-valued function $f\left(\mathbf{X}\right)$ on $\operatorname{UM}\left(N, M\right)$. The Riemannian gradient $\operatorname{grad}f\left(\mathbf{X}\right)$ of $f\left(\mathbf{X}\right)$ at $\mathbf{X}$ represents the steepest ascent denotes as
\begin{equation}
    \operatorname{grad}f\left(\mathbf{X}\right) = \operatorname{P}_{\mathbf{X}}\nabla f\left(\mathbf{X}\right),
\end{equation}
and the specific calculation expression is shown in \eqref{[projection-gradient]}.

{\it Retraction:} On the oblique manifold $\mathcal{OB}\left(2, 2N\right)$, we often use a retraction as follows
\begin{equation}\label{oblique_retraction}
R_{\tilde{\mathbf{Y}}}(\tilde{\xi})\!=\!\left(\mathbf{Y}+\tilde{\xi}\right)\left(\operatorname{ddiag}\left(\left(\mathbf{Y}+\tilde{\xi}\right)^T\left(\mathbf{Y}+\tilde{\xi}\right)\right)\right)^{-1 / 2},
\end{equation}

\begin{lemma}\label{UM_retraction}
    The retraction on $\operatorname{UM}\left(N, 1\right)$ is given by
    \begin{equation}
    \begin{aligned}
       R_{\mathbf{x}}(\xi) &= \phi\left(R_{\tilde{\mathbf{X}}}(\tilde{\xi})\right) \\
       & = \left(\operatorname{ddiag}\left(\left(\mathbf{x}+\xi\right)\left(\mathbf{x}+\xi\right)^H\right) \right)^{-1 / 2} \left(\mathbf{x}+\xi\right).
       \end{aligned}
    \end{equation}
\ \ Then the retraction on $\operatorname{UM}\left(N, M\right)$ is given by
\begin{equation}\label{UM_retraction_operator}
R_{\mathbf{X}}\left(\boldsymbol{\xi}_{\mathbf{X}}\right)=\left(\mathbf{X}+\boldsymbol{\xi}_{\mathbf{X}}\right) \circ \frac{1}{\left|\mathbf{X}+\boldsymbol{\xi}_{\mathbf{X}}\right|_{\left(n,m\right)}},
\end{equation}
\end{lemma}
where
\begin{equation}
\begin{aligned}
&\frac{1}{\left|\mathbf{X}+\boldsymbol{\xi}_{\mathbf{X}}\right|_{\left(n,m\right)}}\\
&\!=\!\!\left[\!\begin{array}{cccc}
\frac{1}{\left|x_1(1)+\xi_{11}\right|}\! &\! \frac{1}{\left|x_2(1)+\xi_{12}\right|} & \cdots & \frac{1}{\left|x_M(1)+\xi_{1M}\right|} \\
\frac{1}{\left|x_1(2)+\xi_{21}\right|} \!&\! \frac{1}{\left|x_2(2)+\xi_{22}\right|} & \cdots & \frac{1}{\left|x_M(2)+\xi_{2M}\right|} \\
\vdots & \vdots & \ddots & \vdots \\
\frac{1}{\left|x_1(N)+\xi_{N1}\right|}\! &\! \frac{1}{\left|x_2(N)+\xi_{N2}\right|} & \cdots & \frac{1}{\left|x_M(N)+\xi_{NM}\right|}
\end{array}\!\right]\!.
\end{aligned}
\end{equation}

{ \it \bf Remark:} This operator represents the matrix obtained by dividing each element of matrix $\mathbf{X}+\boldsymbol{\xi}_{\mathbf{X}}$ by its own modulus (element-wise modulus normalization).

\section{Customized Solving Algorithm}
\subsection{Format Derivation of the Objective Function}
In order to facilitate subsequent derivations, we define the following quantities
\begin{equation}\label{def A}
\begin{split}
\mathbf{A}_\theta = \mathbf{a}_{\theta}\mathbf{a}_\theta^H, \ \ \mathbf{A}_{ij}=\mathbf{a}_{\theta_i} \mathbf{a}_{\theta_j}^H.
\end{split}
\end{equation}
Based on the partial derivative rule of nonsmooth functions of complex variables, we can get
\begin{equation}\label{grad g}
    \frac{\partial g\left(\mathbf{X}\right)}{\mathbf{X}^*} = \sum_{\theta\in\hat{\Theta}}\mathbf{X}\mathbf{A}_\theta -\sum_{\theta\in\bar{\Theta}\cup \theta_c}\mathbf{X}\mathbf{A}_\theta ,
\end{equation}
\begin{equation}\label{grad h}
\begin{aligned}
\frac{\partial\left|P_{\theta_{i,j}, \tau}\right|}{\partial \mathbf{X}^*}&=\frac{\partial\left|\operatorname{Tr}\left(\mathbf{X}^H \mathbf{S}_{\tau} \mathbf{X}  \mathbf{A}_{ji}\right)\right|}{\partial \operatorname{Tr}\left(\mathbf{X}^H \mathbf{S}_{\tau} \mathbf{X}  \mathbf{A}_{ji}\right)} \frac{\partial \operatorname{Tr}\left(\mathbf{X}^H \mathbf{S}_{\tau} \mathbf{X}  \mathbf{A}_{ji}\right)}{\partial \mathbf{X}^*} \\
&\hspace{-1cm}+\frac{\partial\left|\operatorname{Tr}\left(\mathbf{X}^H \mathbf{S}_{\tau} \mathbf{X}  \mathbf{A}_{ij}\right)\right|}{\partial \left(\operatorname{Tr}\left(\mathbf{X}^H \mathbf{S}_{\tau} \mathbf{X}  \mathbf{A}_{ij}\right)\right)^*} \frac{\partial \left(\operatorname{Tr}\left(\mathbf{X}^H \mathbf{S}_{\tau} \mathbf{X}  \mathbf{A}_{ij}\right)\right)^*}{\partial \mathbf{X}^*}\\
& = \frac{z^*}{|z|}\mathbf{S}_{\tau}\mathbf{X} \mathbf{A}_{ji} + \frac{z}{|z|}\mathbf{S}_{\tau}^H\mathbf{X} \mathbf{A}_{ij},
\end{aligned}
\end{equation}
where $z =\operatorname{Tr}\left(\mathbf{X}^H \mathbf{S}_{\tau} \mathbf{X}  \mathbf{A}_{ji}\right)$.

\subsection{Nonsmooth Unimodulus Mnifold Gradient Descend Algorithm}
Let $f: \operatorname{UM}\left(N, M\right) \rightarrow \mathbb{R}$ be a nonsmooth function on a Riemannian manifold. For $\mathbf{X} \in \operatorname{UM}\left(N, M\right)$, we let $\hat{f}_\mathbf{X}=f \circ R_\mathbf{X}$. The Clarke generalized directional derivative of $f$ at $\mathbf{X}$ in the direction $\mathbf{p} \in T_{\mathbf{X}} \operatorname{UM}\left(N, M\right)$, denoted by $f^{\circ}(\mathbf{X}; \mathbf{p})$, is defined by $f^{\circ}(\mathbf{X}; \mathbf{p})=\hat{f}_x^{\circ}\left(\mathbf{0}_\mathbf{X}; \mathbf{p}\right)$, where $\hat{f}_\mathbf{X}^{\circ}\left(\mathbf{0}_\mathbf{X}; \mathbf{p}\right)$ denotes the Clarke generalized directional derivative of $\hat{f}_\mathbf{X}: T_\mathbf{X} \operatorname{UM}\left(N, M\right) \rightarrow \mathbb{R}$ at $\mathbf{0}_\mathbf{x}$ in the direction $\mathbf{p} \in T_\mathbf{X} \operatorname{UM}\left(N, M\right)$. Therefore, the generalized subdifferential of $f$ at $\mathbf{X}$, denoted by $\partial f(\mathbf{X})$, is defined by $\partial f(\mathbf{X})=\partial \hat{f}_\mathbf{X}\left(\mathbf{0}_\mathbf{X}\right)$.
Then, we use $\varepsilon$-subdifferential to approximate the calculation of subdifferential.

Let $f: \operatorname{UM}\left(N, M\right) \rightarrow \mathbb{R}$ be a nonsmooth function on a Riemannian manifold $\operatorname{UM}\left(N, M\right)$. We define the $\varepsilon$-subdifferential of $f$ at $\mathbf{X}$ denoted by $\partial_{\varepsilon} f(\mathbf{x})$ as follows;
\begin{equation}
\begin{aligned}
   \hspace{-0.2cm} \partial_{\varepsilon} f(\mathbf{X})=&\operatorname{clconv}\!\left\{\beta_\eta^{-1} \mathcal{T}_{\mathbf{X} \leftarrow \hat{\mathbf{\mathbf{X}}}}(\partial f(\hat{\mathbf{\mathbf{X}}}))\!:\right. \\
   \hspace{-0.2cm}  & \left. \!\hat{\mathbf{\mathbf{X}}}\! \in \!\operatorname{cl} B_R(\mathbf{X}, \varepsilon),  \text { and } \eta=R_\mathbf{X}^{-1}(\hat{\mathbf{X}})\right\},
    \end{aligned}
\end{equation}
where $\mathcal{T}_{\mathbf{X} \leftarrow \hat{\mathbf{\mathbf{X}}}}(\partial f(\hat{\mathbf{\mathbf{X}}})) = \operatorname{P}_{\mathbf{X}}\partial f\left(\hat{\mathbf{X}}\right)$ and $\beta_{\eta} = \frac{\|\eta\|}{\|\mathcal{T}_{R_{\eta}}\eta\|}$.
Every element of the $\varepsilon$-subdifferential is called an $\varepsilon$-subgradient.
The proposed nonsmooth unimodulus manifold gradient descend (N-UMGD) algorithm is presented in Table \ref{table ADMM}.

\begin{table}[t]
\caption{The proposed N-UMGD algorithm }
\label{table ADMM}
\centering
\begin{tabular}{l}
\hline\hline
{\bf Initialization:}  \\
 \hspace{0.2cm} Set iteration index $k\!=\!0$, initialize $\mathbf{X} \in \operatorname{UM}\left(N, M\right)$ ,\\
 \hspace{0.2cm} and $W_k:=\left\{\mathbf{v}_1, \ldots, \mathbf{v}_k\right\} \subseteq \partial_{\varepsilon} f(\mathbf{X})$ randomly,\\
 {\bf repeat} \\
 \hspace{0.2cm} S.1 Choose a descent direction $\mathbf{p}_k \in T_{\mathbf{X}_k}\mathcal{M}$ satisfies\\
\hspace{2.7cm} $\underset{\mathbf{p}_k \in \operatorname{conv} W_k}{\arg \min } \|\mathbf{p}_k\|$,\\
 \hspace{0.2cm} S.2 Choose a step length $\alpha_k \in \mathbb{R}$,\\
 \hspace{0.2cm} S.3 update $\mathbf{X}_k = R_{\mathbf{X}_k}\left(-\alpha_k\mathbf{p}_k\right)$,\\
{\bf until} some preset termination criterion is satisfied.\\
\hspace{0.75cm} Let $\mathbf{X}^{k+1}$ be the output.\\
\hline\hline
\end{tabular}
\end{table}

\subsection{Algorithm Analysis}
\subsubsection{Implementation Aanalysis}
The computational complexity of the proposed N-UMGD algorithm is dominated by  the gradient of objective function.
\begin{itemize}
    \item For $g(\mathbf{X})$, constant matrix $\mathbf{A_\theta}$ from different $\theta$ can be calculated only once and combined as follows
    $$   \bar{\mathbf{A}}_\theta = \displaystyle\sum_{\theta\in\hat{\Theta}}\mathbf{A}_\theta -\displaystyle\sum_{\theta\in\bar{\Theta}\cup \theta_c}\mathbf{A}_\theta.$$

    Therefore, the computational complexity of gradient function $\nabla g(\mathbf{X})$ is $\mathcal{O}(M^2N)$ according to \eqref{grad g}.
    \item For  $h(\mathbf{X})$, the computational complexity of $\frac{\partial\left|P_{\theta_{i,j}, \tau}\right|}{\partial \mathbf{X}^*}$ is $\mathcal{O}(2M^2N)$ according to \eqref{grad h}. The computational complexity of gradient function $\nabla h(\mathbf{X})$ is $\mathcal{O}(2K^2|\Gamma|M^2N)$, where $|\Gamma|$ is the size of $\Gamma$.
\end{itemize}

In summary, the total cost of the proposed N-UMGD algorithm in each iteration is roughly $\mathcal{O}(2K^2|\Gamma|M^2N)$.

\subsubsection{Convergence Issue}
 The proposed  N-UMGD algorithm is terminable and convergent if proper parameters are chosen.
 We have {\it Theorem \ref{theorem}} to show its convergence characteristic.

\begin{theorem}\label{theorem}
    $\operatorname{UM}\left(N,M\right)$ is a Riemannian manifold. For $\mathbf{X} \in \operatorname{UM}\left(N,M\right)$, $f\left(\mathbf{X}\right) = g\left(\mathbf{X}\right)+h\left(\mathbf{X}\right)$ is a Lipschitz function of Lipschitz constant $L$ near $\mathbf{X}$, i.e., $|f\left(\mathbf{X}\right)-f\left(\mathbf{Y}\right)| \leq L\operatorname{dist}\left(\mathbf{X}, \mathbf{Y}\right)$ for all $\mathbf{Y}$ in a neighborhood of $\mathbf{X}$. Then, if $\{\mathbf{X}_k\}$ and $\{\mathbf{\xi}_k\}$ are sequences in $\operatorname{UM}\left(N,M\right)$ and $T\operatorname{UM}\left(N,M\right)$ such $\mathbf{\xi}_k \in \partial f\left(\mathbf{X}_k\right)$ for each $i$, and if $\{\mathbf{X}_k\}$ converges to $\mathbf{X}$ and $\mathbf{\xi}$ is a cluster point of the sequence $\{\mathbf{\xi}_k\}$, then every accumulation point of the sequence $\{\mathbf{X}_k\}$ belongs to the set $\{\mathbf{X} \in \operatorname{UM}\left(N,M\right): 0 \in \partial f\left(\mathbf{X}\right)\}$, that means $\mathbf{X}^*$ is an $\epsilon-$stationary point.\footnote{Due to limited space,  the detailed proofs of the lemmas and theorems mentioned in this paper are not included, and will be presented in future work.}
\end{theorem}

\section{Simulation results}\label{simulation}
\begin{figure}[t]
\centerline{\includegraphics[scale=0.6]{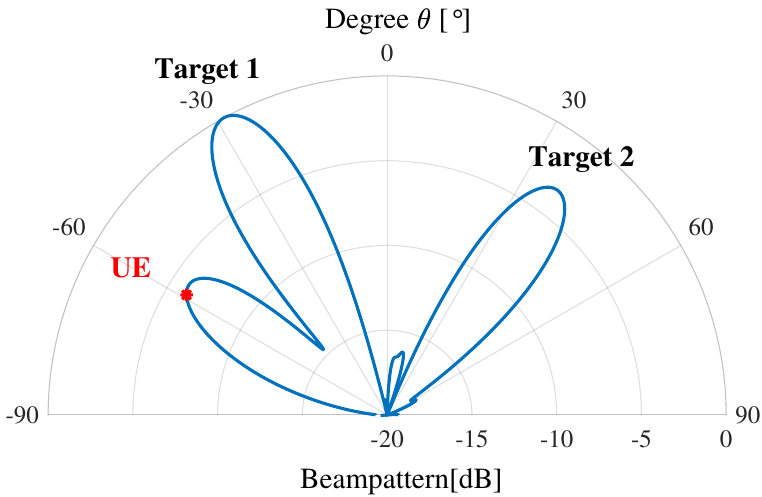}}
\caption{Synthesized beampattern.}
\label{beampattern}
\end{figure}
\begin{figure}[t]
\centerline{\includegraphics[scale=0.43]{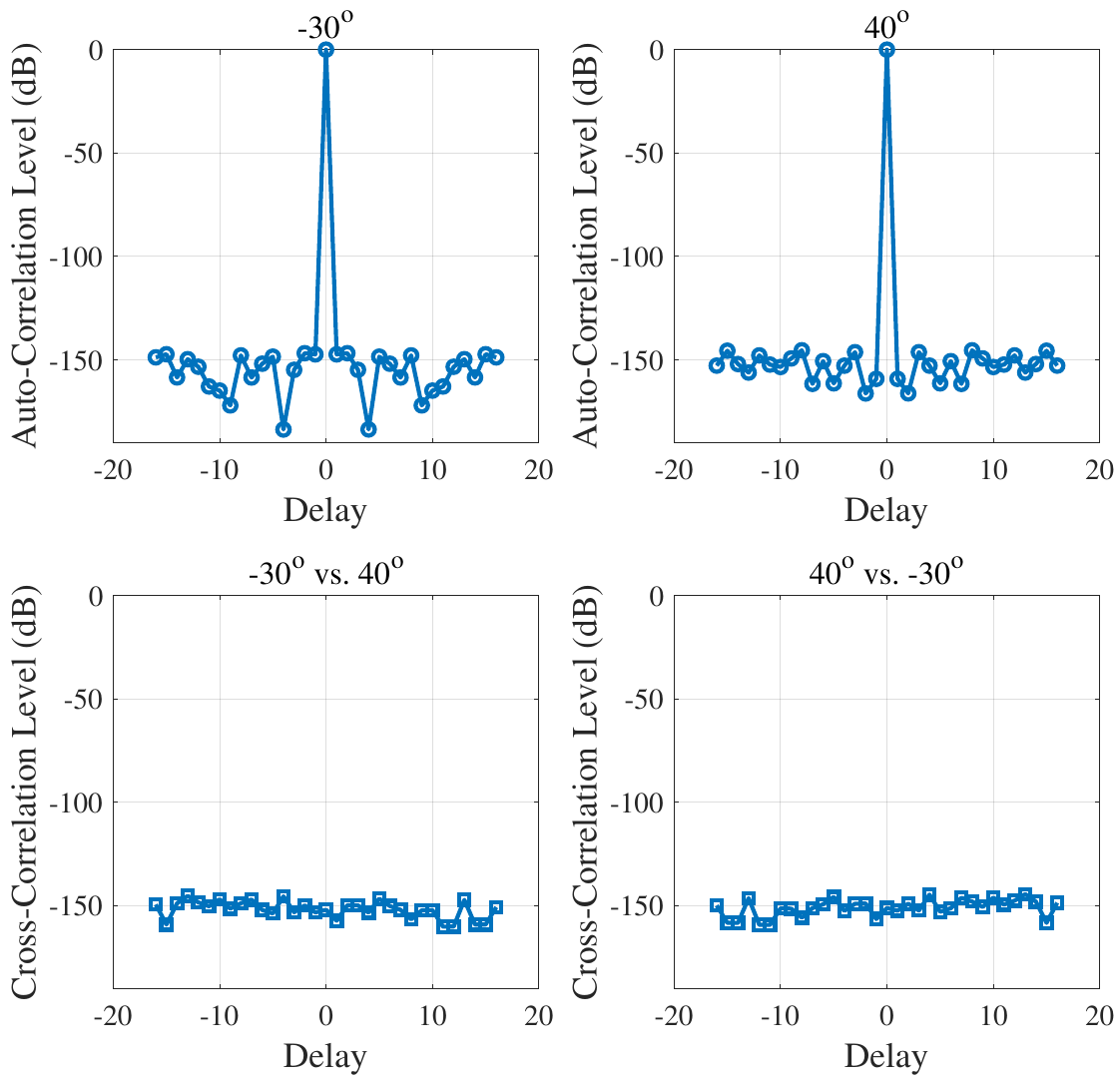}}
\caption{Correlation characteristics.}
\label{correlation}
\end{figure}

\begin{figure}[htbp!]
\centerline{\includegraphics[scale=0.6]{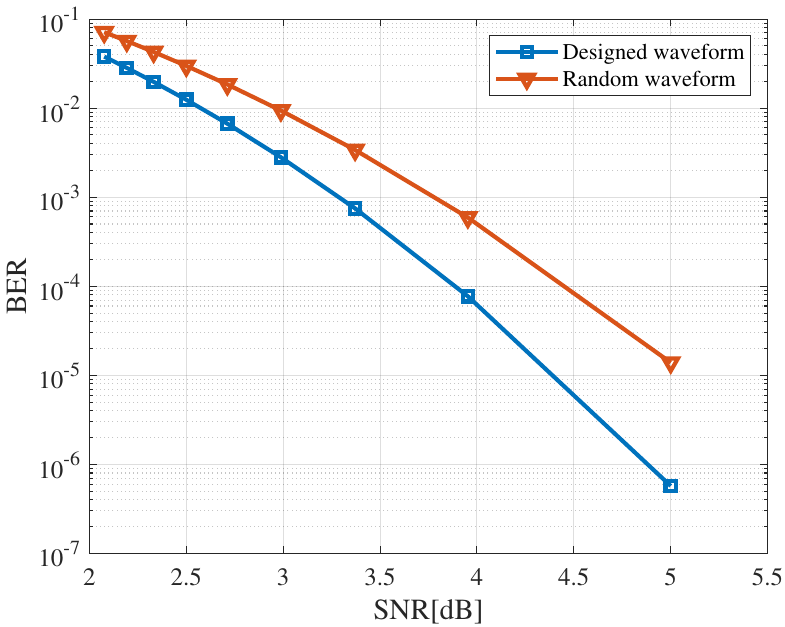}}
\caption{BER performance.}
\label{ber}
\end{figure}

This section presents numerical results to demonstrate the performance of the designed waveform generated by the proposed N-UMGD algorithm for ISAC system.
Without loss of generality, we assume that the array spacing is half wavelength, and the set of spacial angles covers $(-90^\circ,90^\circ)$ with spacing $0.1^\circ$.
The number of antennas is $M = 8$  and waveform length is $N =128$.
The spacial delay internal is $[0,16]$.
We consider two target directions $\theta_1=-30^\circ$ and $\theta_2=40^\circ$, and the communication user at $\theta_c = -60^\circ$ direction.

\subsection{Radar Sensing Performance}
From Fig. \ref{beampattern}, we can see that
the signal energy is mainly concentrated in two desired target directions and the communication user direction.
In this way, the dual functions of the ISAC system can be realized, that is, good radar sensing performance can be achieved, and information transmission can be realized.

Fig. \ref{correlation} shows that the normalized spacial auto/cross-correlation label can be as low as about $-150$dB.
 By obtaining a lower spatial correlation level, the clutter component can be effectively suppressed to achieve better radar sensing performance.
 Compared to classical $\ell_2$-norm model, the results in Fig. \ref{correlation} verify the superiority of the $\ell_1$-norm model proposed in this paper.

\subsection{Communication Performance}
Fig. \ref{ber} shows that compared with random waveform, the waveform designed  by the proposed N-UMGD algorithm can not only realize the communication function, but also enjoy better BER performance (roughly $0.5$dB).
This shows the effectiveness and superiority of the customized model and the proposed algorithm designed in this paper.
It is believed that with the increase of waveform dimension, both radar sensing and communication functions will have better performance.

\section{Conclusion}

In this paper, we formulate the waveform design problem of ISAC system as a nonsmooth and nonconvex model with unimodulus constraint.
By analyzing the inherent geometric structure of unimodulus constraint, the problem is transformed into an unconstrained problem on Riemannian manifold, and then the customized N-UMGD algorithm is proposed by constructing operators on Riemannian space.
The computational complexity and convergence of the proposed algorithm are analyzed.
Finally, the simulation results demonstrate the effectiveness of the designed waveform, that is, it can achieve good radar sensing and communication functions.

\ifCLASSOPTIONcaptionsoff
  \newpage
\fi


\begin{thebibliography}{99}

\bibitem{Shlezinger_20}
D. Ma, N. Shlezinger, T. Huang, Y. Liu and Y. C. Eldar, ``Joint RadarCommunication Strategies for Autonomous Vehicles: Combining Two
Key Automotive Technologies,''  \emph{IEEE Signal Processing Magazine}, vol. 37, no. 4, pp. 85-97, July 2020.

\bibitem{Zhou_22}
W. Zhou, R. Zhang, G. Chen and W. Wu,``Integrated Sensing and Communication Waveform Design: A Survey,'' \emph{IEEE Open Journal of the Communications Society}, vol. 3, pp. 1930-1949, 2022.
\bibitem{Zhang_21}
J. A. Zhang, et al., ``An Overview of Signal Processing Techniques for Joint Communication and Radar Sensing,'' \emph{IEEE Journal of Selected Topics in Signal Processing}, vol. 15, no. 6, pp. 1295-1315, Nov. 2021.


\bibitem{Hassanien_16}
A. Hassanien, M. G. Amin, Y. D. Zhang and F. Ahmad, ``Signaling strategies for dual-function radar communications: an overview,'' \emph{IEEE Aerospace and Electronic Systems Magazine}, vol. 31, no. 10, pp. 36-45, Oct. 2016.
\bibitem{Liu_18_twc}
F. Liu, C. Masouros, A. Li, H. Sun and L. Hanzo, ``MU-MIMO Communications With MIMO Radar: From Co-Existence to Joint Transmission,'' \emph{IEEE Transactions on Wireless Communications}, vol. 17, no. 4, pp. 2755-2770, Apr. 2018.

\bibitem{Shi_18}
C. Shi, F. Wang, M. Sellathurai, J. Zhou and S. Salous, ``Power Minimization-Based Robust OFDM Radar Waveform Design for Radar and Communication Systems in Coexistence,''  \emph{IEEE Transactions on Signal Processing}, vol. 66, no. 5, pp. 1316-1330, Mar. 2018.


\bibitem{Liu_18}
F. Liu, L. Zhou, C. Masouros, A. Li, W. Luo, and A. Petropulu, ``Toward dual-functional radar-communication systems: Optimal waveform design,''  \emph{IEEE Trans. Signal Process.}, vol. 66, no. 16, pp. 4264-4279, Aug. 2018.

\bibitem{Cao_20}
N. Cao, Y. Chen, X. Gu and W. Feng, ``Joint Radar-Communication Waveform Designs Using Signals From Multiplexed Users,'' \emph{IEEE Transactions on Communications}, vol. 68, no. 8, pp. 5216-5227, Aug. 2020.

\bibitem{Huang_22}
Y. Huang, S. Hu, S. Ma, Z. Liu and M. Xiao, ``Designing Low-PAPR Waveform for OFDM-Based RadCom Systems,'' \emph{IEEE Transactions on Wireless Communications}, vol. 21, no. 9, pp. 6979-6993, Sept. 2022.

\bibitem{He_24}
Y. He, G. Yu, Z. Tang, J. Wang and H. Luo, ``A Dual-Functional Sensing-Communication Waveform Design Based on OFDM,'' \emph{IEEE Transactions on Wireless Communications}, doi: 10.1109/TWC.2024.3448456.

\bibitem{Keskin_21}
M. F. Keskin, V. Koivunen and H. Wymeersch, ``Limited Feedforward Waveform Design for OFDM Dual-Functional Radar-Communications,'' \emph{IEEE Transactions on Signal Processing}, vol. 69, pp. 2955-2970, 2021.

\bibitem{Bazzi_23}
A. Bazzi and M. Chafii, ``On Integrated Sensing and Communication Waveforms With Tunable PAPR,'' \emph{IEEE Transactions on Wireless Communications}, vol. 22, no. 11, pp. 7345-7360, Nov. 2023.



\bibitem{Liu_21_jstsp}
R. Liu, M. Li, Q. Liu and A. L. Swindlehurst, ``Dual-Functional Radar-Communication Waveform Design: A Symbol-Level Precoding Approach,'' \emph{IEEE Journal of Selected Topics in Signal Processing}, vol. 15, no. 6, pp. 1316-1331, Nov. 2021.

\bibitem{Xiao_22}
Z. Xiao and Y. Zeng, ``Waveform Design and Performance Analysis for Full-Duplex Integrated Sensing and Communication,'' \emph{IEEE Journal on Selected Areas in Communications}, vol. 40, no. 6, pp. 1823-1837, June 2022.

\bibitem{Zhang_24}
F. Zhang, T. Mao, R. Liu, Z. Han, S. Chen and Z. Wang, ``Cross-Domain Dual-Functional OFDM Waveform Design for Accurate Sensing/Positioning,'' \emph{IEEE Journal on Selected Areas in Communications}, vol. 42, no. 9, pp. 2259-2274, Sept. 2024.

\bibitem{Hassanien_16_tsp}
A. Hassanien, M. G. Amin, Y. D. Zhang and F. Ahmad, ``Dual-Function Radar-Communications: Information Embedding Using Sidelobe Control and Waveform Diversity,'' \emph{IEEE Transactions on Signal Processing}, vol. 64, no. 8, pp. 2168-2181, Apr. 2016.

\bibitem{Kumari_20}
P. Kumari, S. A. Vorobyov and R. W. Heath, ``Adaptive Virtual Waveform Design for Millimeter-Wave Joint Communication-Radar,'' \emph{IEEE Transactions on Signal Processing}, vol. 68, pp. 715-730, 2020.









\bibitem{Statistical_signal_processing}
S. M. Kay, \emph{Fundamentals of Statistical Signal Processing: Estimation Theory}. Prentice Hall, 1993.

\bibitem{Oblique_manifold}
P. A. Absil and K. A. Gallivan, ``Joint Diagonalization on the Oblique Manifold for Independent Component Analysis,'' \emph{in 2006 IEEE International Conference on Acoustics Speech and Signal Processing Proceedings}, Toulouse, France, 2006.

\bibitem{manifold-nicolas}
Boumal N. \emph{An introduction to optimization on smooth manifolds}. Cambridge University Press, 2023.



%
%
%
%
%
%
%
%
%
%
%
%
%
%
%
%
%
%
%
%
%
%
%
%
%
%
%
%
%
%
%
%
%
%
%
%

\end{thebibliography}
\end{document}